\documentclass[reprint,amsmath,amssymb,aps,prx,tightenlines,nobibnotes,superscriptaddress,longbibliography]{revtex4-2}
\usepackage{centernot}
\usepackage{graphicx}
\usepackage{amsmath}
\usepackage{times}
\usepackage{amssymb}
\usepackage{mathrsfs}
\usepackage{chemarr}
\usepackage{xcolor}
\usepackage{version}
\usepackage{enumitem}

\newcommand\cM{\mathcal{M}}
\newcommand\cD{\mathcal{D}}
\newcommand\cQ{\mathcal{Q}}
\newcommand\br{{\bf r}}

\newcommand\bff{{\bf f}}

\newcommand\bq{{\bf q}}
\newcommand\rmd{{\rm d}}

\newcommand\rmT{{\rm T}}
\newcommand\rmKB{{\rm K}}
\newcommand\mN{\mathcal{N}}
\newcommand{\mv}[1]{{ \langle#1\rangle}}

\graphicspath{{./Image/}}
\definecolor{linkcolor}{rgb}{0,0,0.6}
\usepackage{lipsum}
\usepackage{tikz}

\begin{document}

\title{Orientational order on non-orientable domains}

\author{Gianmarco Spera}
\affiliation{Rudolf Peierls Centre for Theoretical Physics, University of Oxford, Oxford OX1 3PU, United Kingdom}
\author{Axel Fotso Ndefo}
\affiliation{Rudolf Peierls Centre for Theoretical Physics, University of Oxford, Oxford OX1 3PU, United Kingdom}

\author{Keaton J. Burns}
\affiliation{Department of Mathematics, Massachusetts Institute of Technology, Cambridge, MA 02139}
\affiliation{Center for Computational Astrophysics, Flatiron Institute, New York, NY 10010}

\author{Alexander Mietke}
\email{alexander.mietke@physics.ox.ac.uk}
\affiliation{Rudolf Peierls Centre for Theoretical Physics, University of Oxford, Oxford OX1 3PU, United Kingdom}


\begin{abstract}
We study the statistical properties of passive and active many-body systems with orientational degrees of freedom on non-orientable domains. By rephrasing topological constraints as non-local symmetry relations on an orientable double-cover, we show that non-orientability eliminates global rotational soft modes without acting like an external field. In a passive XY model, this results in topological caging, where orientational fluctuations that exhibit conventional diffusive behavior on a torus saturate on a Klein bottle to a finite value that we compute exactly in the thermodynamic limit. In models of active self-propelled particles with orientational degrees of freedom, topological caging persists despite continuously changing interaction neighborhoods. In an active Ising spin model, non-orientability enforces the coexistence of ordered anti-parallel domains with vanishing global polar order, a state that is absent on orientable domains.
\end{abstract}

\maketitle
Collective properties of physical systems are inherently linked to the topology of the domain on which they evolve~\cite{gold12}. In systems with orientational degrees of freedom the competition between topological constraints and the preference for global orientational order leads to frustration and the emergence of non-trivial patterns~\cite{isham1989modern}. A well-known example are nematic liquid crystals that develop topological defects when constrained to spherical surfaces~\cite{vitelli2006nematic}. Similarly frustrated states can be observed in systems with polar order parameters~\cite{baek2009curvature,selinger2011monte}. Topological frustration also impacts active systems that exhibit flocking-like dynamics~\cite{shankar2017topological,shankar2022topological,hueschen2023wildebeest,tan24} and it contributes to the robustness of symmetry breaking processes during organism development~\cite{maroudas2021topological,ravi25}. 

A topological property whose impact on systems with orientational degrees of freedom is less well studied is domain \textit{orientability}. A spherical surface is orientable because an orientation in terms of a smoothly varying surface normal can be defined globally. M\"obius strips and Klein bottles are canonical examples of \textit{non-orientable} domains~\cite{spivak2018calculus,braselton2002surface}, on which local surface normals flip their direction when transported along a closed paths. Non-orientability thereby imposes constraints on excitable modes in both quantum and classical systems, giving rise to unconventional optical, acoustic, electronic and phononic effects~\hbox{\cite{zhao2009observable,jiang2010topological,beugeling2014nontrivial,nishiguchi2018phonon,blio21,beli21,chen22,flouris2022curvature}}. The response of liquid crystals on embedded twisted strips has been studied numerically in great detail~\cite{mach13}. Further work supported by experiments has rationalized the collapse of non-orientable fluid films~\cite{gold10,gold14,moff16}, and found that solid M\"obius ribbons serve as efficient actuators~\cite{nie2021light}, exhibit non-reciprocal elasticity~\cite{bartolo2019topological}, and act as topologically protected memory~\cite{guo2023non,moguel2025topological}. So far, such studies have been restricted to deterministic scenarios.

Without global orientation there is no notion of an arbitrary homogeneous in-plane rotation, which is -- by contrast -- an accessible soft mode in ferromagnetic and liquid crystal theories on closed orientable surfaces~\cite{goldenfeld2018lectures}. In this {\it Letter}, we use stochastic microscopic models and mean field theory to explore how such topological mode suppression impacts the statistical properties of many-particle systems with ferromagnetic interactions. To this end, we rephrase the topological constraints of a Klein-bottle in terms of non-local constraints on a flat orientable double-cover. This enables exact analytic calculations of correlations in finite systems and in the thermodynamic limit (TDL), as well as efficient numerical simulations without the need to describe curved embedded surfaces. We show that orientational fluctuations of an~XY spin lattice become bounded on non-orientable domains, which we refer to as \textit{topological caging}. We derive exact analytic results to explain this observation and discuss the analogy of this topological effect with the symmetry-breaking due to an external field. Finally, we show how non-orientability suppresses emerging orientational~order in microscopic flocking~models. 


\begin{figure*}[t!]
    \centering
    \includegraphics[width = 2.05\columnwidth]{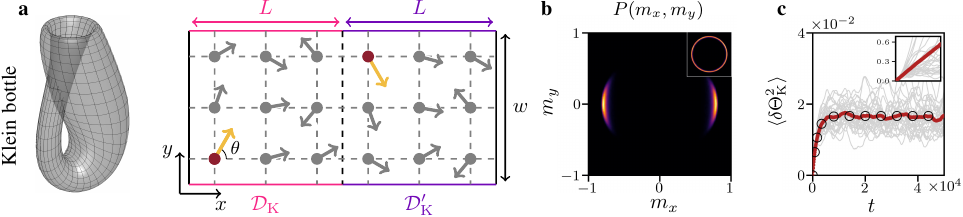}\vspace{-0.2cm}
    \caption{
    \textbf{Double-cover representation and fluctuations of the XY model on a Klein bottle.} \textbf{(a)}~Double-cover $\Omega_K=\cD_K\cup\cD'_K$ of a non-orientable Klein bottle surface of size $L\times w$ with twist-axis $x$ and periodicity in $y$-direction: $\cD'_K$ represents an observation of $\cD_K$ from the backside. Points are identified as $(x,y)\sim(x+L,w-y)$. Spin components orthogonal to the twist axis transform as pseudoscalars, \hbox{$f_y^{\rmKB}(x+L, y) = -f_y^{\rmKB}(x, w-y)$} (see colored spin pair). \textbf{(b)}~In an XY lattice model on a Klein bottle surface, topological pinning of spin orientations along the twist-axis leads to a collapse of the polarization distribution $P(m_x,m_y)$ onto $m_x\approx\pm1,\,m_y\approx0$. Inset: $P(m_x,m_y)$ is isotropic on a torus (see Methods). \textbf{(c)}~Topological caging: Orientation fluctuations $\langle \delta \Theta^2(t) \rangle = \langle [\Theta(t+t_0)-\Theta(t_0)]^2)$ [see Eq.~(\ref{eq:defM})] on a Klein bottle are bound [gray: individual realizations, red: average, black circles: analytic prediction Eq.~(\ref{eq:ThetaKB_t})]. Inset: Orientational fluctuations $\langle \delta \Theta^2_T\rangle$ exhibit conventional diffusion on a torus~\cite{supp}. Parameters: See~Methods.
    }
    \label{fig:fig1}
\end{figure*}

{\it Non-orientability with flat embedding in 2D.} The topological constraints of a Klein bottle surface (Fig.~\ref{fig:fig1}a) can be rephrased as non-local constraints on an orientable double-cover~\cite{gold10,bala26}. While the double-cover construction is trivial for a domain of size $L\times w$ with toroidal topology~(see Methods), a M\"obius strip of the same size -- which becomes a Klein-bottle domain~$\mathcal{D}_K$ when opposite strip boundaries are periodically glued together -- is in the direction of its \textit{twist-axis} only fully periodic along paths of length~$2L$ (Fig.~\ref{fig:fig1}a). One can then interpret the second cover~$\mathcal{D}_K'$ as an observation of~$\mathcal{D}_K$ from the backside. Accordingly, we must identify points $(x,y)\in\mathcal{D}_K$ with points \hbox{$(x+L,w-y)\in\mathcal{D}_K'$}, where we use a common coordinate system across the double-cover and align the $x$-direction with the twist-axis. Vector fields ${\bf f}=(f_x,f_y)$ on a torus trivially satisfy ${\bff}^{\rmT}(x+L, y+w) = \bff^{\rmT}(x,y)$. On a Klein bottle, vector components along the twist-axis transform under translations as scalars, \hbox{$f_x^{\rmKB}(x+L, y) = f_x^{\rmKB}(x, w-y)$}, components orthogonal to the twist axis transform as pseudoscalars, \hbox{$f_y^{\rmKB}(x+L, y) = -f_y^{\rmKB}(x, w-y)$} (see colored spin in Fig.~\ref{fig:fig1}a). Importantly, this formulation does not create a physical boundary along the twisted $x$-direction. Instead, all vector fields must satisfy -- in addition to the $2L$-periodicity of the double-cover ${\bff}^{\rmKB}(x+2L, y+w) = \bff^{\rmKB}(x,y)$ -- an \hbox{$L$-periodicity} along the lines $y=0$ and $y=w/2$. A crucial implication is that \textit{spatially constant} pseudoscalar vector components $f_y^K=f_0$ would have to satisfy $f_0=-f_0$ at two different points of the double-cover and must therefore vanish. To make this more intuitive, consider an orientation field $\theta(x,y)$ measured relative to the $x$-axis (see Fig.~\ref{fig:fig1}a). On a Klein bottle double-cover, vector component transformation rules imply for such a field
\begin{equation}\label{eq:angmod}
\theta^K(x+L,y)=-\theta^K(x,w-y)\mod 2\pi,    
\end{equation}
where we use that global angular fields are well-defined on the orientable double-cover ${\Omega}_K = \cD_K\cup \cD_K'$. For constant $\theta^K(x,y)=\theta_0$, Eq.~(\ref{eq:angmod}) implies $\theta_0\in\{0,\pi\}$, i.e. homogeneous orientation fields are forced to align with the twist axis and homogeneous global rotations by arbitrary angles are topologically suppressed. 

{\it XY model on non-orientable domain.}
We first consider a system of $N$ XY spins on a 2D square lattice of size \hbox{$N=L\times w$} with lattice point spacing \hbox{$dx=dy=a=1$}. Spins interact via ferromagnetic alignment and their orientations $\theta_i$ evolve~as 
\begin{equation}\label{eq:xy-dynamics}
    \partial_t \theta_i = \gamma \sum_{j \in \mN_i} \sin(\theta_j-\theta_i) + \sqrt{2D_r} \zeta_i \;,
\end{equation}
where $i=(i_x, i_y)$ denotes a multi-index on the lattice. In Eq.~(\ref{eq:xy-dynamics}), $\gamma$ is the alignment strength, $\mN_i$ is the set of nearest neighbors of particle $i$, $D_r$ is the rotational diffusivity, and $\zeta_i$ is a Gaussian white noise. To impose a Klein bottle domain topology, we simulate in practice $2N$ spins on a $(2L,w)$-grid and impose on the pseudoscalar orientation $\theta_i$ of suitable spin pairs the constraint \hbox{$\theta(i_x+L, i_y) = - \theta(i_x, w - i_y)\,\mod 2\pi$}, in analogy to Eq.~(\ref{eq:angmod}). To characterize the dynamics of this model, we measure the complex order parameter
\begin{equation}\label{eq:defM}
\cM = |\mathcal{M}| {\rm e}^{i \Theta}=\frac{1}{N} \sum_{i \in \cD} {\rm e}^{i\theta_i}\ , 
\end{equation}
over $N$ independent spins. The sum in Eq.~(\ref{eq:defM}) defines the polar order parameter $|\mathcal{M}|$, as well as the global orientation \smash{$\Theta= {\rm arg}(\cM)$}. Real and imaginary parts of $\cM$, \smash{$m_x  = N^{-1} \sum_{i \in \cD} \cos\theta_i$} and \smash{$m_y  = N^{-1} \sum_{i \in \cD} \sin\theta_i$}, respectively, quantify the average spin polarization along $x$ and~$y$. 


\begin{figure*}[ht!]
    \centering
   \includegraphics[width = 2.05\columnwidth]{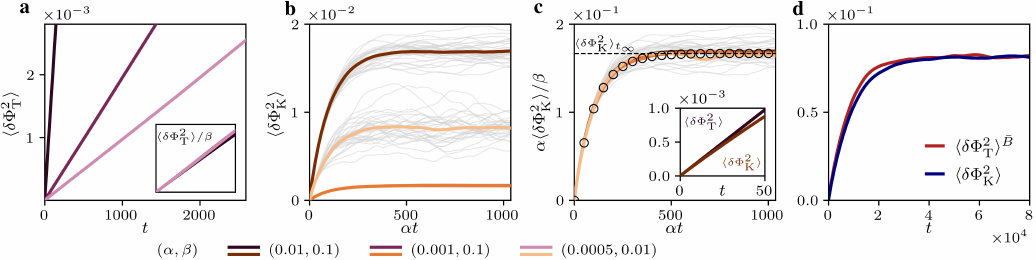}
    \caption{\textbf{Mean-field theory of orientation fluctuations on torus and Klein bottle.} \textbf{(a)}~Spin-wave approximation of XY model, Eq.~(\ref{eq:meanfield-xy}), on torus yields expected diffusion of average orientations $\langle \delta\Phi^2_{\rm T}\rangle$ [see Eqs.~(\ref{eq:global-theta}),(\ref{eq:ThetaT_t})] that only depends on noise strength $\beta$~(inset). \textbf{(b)}~Fluctuations in the mean-field theory on a Klein bottle reproduce bound fluctuations of lattice model (see Fig.~\ref{fig:fig1}f). \textbf{(c)}~Data from panel (b) agrees with exact analytic prediction [black circles, see Eq.~(\ref{eq:ThetaKB_t})] and collapses under corresponding scaling. Black dashed line indicated fluctuation plateau, Eq.~\eqref{eq:tdlimitKB}, in the thermodynamic limit. Inset illustrates analytically predicted identical fluctuation dynamics at short times between torus and Klein bottle. \textbf{(d)}~Comparison between fluctuations on torus \smash{$\langle\Phi_T^2\rangle^{\bar{B}}$} with an external field $\bar{B}$ tuned such that plateau-value agrees with topologically induced bound on a Klein bottle. Parameters: See Methods.
    }
    \label{fig:fig2}
\end{figure*}

On a domain with toroidal topology, Eq.~(\ref{eq:angmod}) results in the well-known ring-like spin-polarization distribution~$P(m_x, m_y)$~\cite{goldenfeld2018lectures} (Fig.~\ref{fig:fig1}b, inset) and diffusive behavior of fluctuations $\langle \delta \Theta^2(t) \rangle = \langle [\Theta(t+t_0)-\Theta(t_0)]^2 \rangle$ of the global unwinded orientation~$\Theta$~(Fig.~\ref{fig:fig1}c, inset, see Methods and \cite{supp}). Numerical simulations of the same model on a domain with Klein bottle topology reveal instead an equilibrated spin polarization distribution that peaks around $m_y\approx 0$ and \hbox{$m_x=\pm1$}~(Fig.~\ref{fig:fig1}b), indicating a topological suppression of the diffusive global rotation mode seen on a torus. Furthermore, global orientation fluctuations~$\langle \delta \Theta^2_{\text{K}}\rangle$ exhibit on a Klein bottle a saturation behavior~(Fig.~\ref{fig:fig1}c) that is induced by the topological constraints and reminiscent of particle diffusion in environments where particles get caged~\cite{weeks2002subdiffusion,chepizhko2013diffusion,zeitz2017active,pietrangeli2025universal}.

We can qualitatively rationalize these numerical observations by introducing an energetic description of the model Eq.~\eqref{eq:xy-dynamics} in the form $\partial_t\theta_i=-\delta E/\delta\theta_i+\sqrt{2D_r}\zeta_i$  on the double-cover \hbox{$\Omega=\cD\cup\cD'$}, where
\begin{align}   
E = &\,\frac{\gamma}{2} \sum_{i \in \Omega} \sum_{j \in \mN_i}\left[1- \cos(\theta_j - \theta_i)\right] + \sum_{i \in \cD} \mathcal{T}(\theta_i,\theta_{k})\ ,\label{eq:domain-energy}
\end{align}
and $\mathcal{T}(\theta_i,\theta_{k})$ enforces the topological constraints between spin pairs on lattice sites $i$ and $k$: $\mathcal{T}$ is zero if the constraint is satisfied or the spins are independent, and $\mathcal{T}$ is infinite if the topological constraint is violated. On a Torus, $\mathcal{T}$ enforces \hbox{$\theta(i_x,i_y) - \theta(i_x+L, i_y) = 0$} (dropping$\mod 2\pi$ from here on), such that Eq.~(\ref{eq:domain-energy}) contains the known rotational soft mode $\theta_i=\theta_0\in[0,2\pi)$. On a Klein bottle, we have instead the constraint \hbox{$\theta(i_x,i_y) + \theta(i_x+L, w - i_y)= 0$}, which enforces \smash{$\theta_0^{(1)}=0$} and \smash{$\theta_0^{(2)}=\pi$} to be the only allowed homogeneous orientations. A topologically consistent process to switch between states \smash{$\theta_0^{(1)}$} and \smash{$\theta_0^{(2)}$} thus inevitably incurs an energetic penalty via the first term in Eq.~(\ref{eq:domain-energy}). This topologically induced energy barrier is consistent with the symmetric collapse of the equilibrated spin polarization distribution~(Fig.~\ref{fig:fig1}b) and the reduction of global orientational fluctuations~(Fig.~\ref{fig:fig1}c). 

{\it Stochastic mean-field theory.}
To explain our numerical observations quantitatively, we employ a spin-wave approximation and consider small angle variations $ \varphi = \theta - \theta_0 $ around an ordered state with $\theta_0$. The dynamics~\eqref{eq:xy-dynamics} is then approximated~by
\begin{equation}\label{eq:meanfield-xy}
    \partial_t \varphi(\br, t) = \alpha \Delta \varphi(\br,t) + \sqrt{2 \beta} \xi(\br,t)\;,
\end{equation}
where $\alpha=\gamma a^2$ is the elastic constant, $\beta=D_ra^2$ is the noise strength, and $\xi$ is a Gaussian white noise with unit variance.  Numerical measurements of the average orientation 
\begin{equation}\label{eq:global-theta}
    \Phi(t) = \frac{1}{Lw}\int \rmd\br\,\varphi(\br,t)\,,
\end{equation}
and its fluctuations, $\langle \delta\Phi^2(t) \rangle = \mv{\left[\Phi(t+t_0) - \Phi(t_0)\right]^2}$, on a torus~(Fig.~\ref{fig:fig2}a) and a Klein bottle~(Fig.~\ref{fig:fig2}b) reproduce the two distinct behaviors seen in the lattice model (Fig.~\ref{fig:fig1}c). We then consider a Fourier expansion $\varphi(\br,t) = \sum_{\bq \in \cQ} \tilde \varphi(\bq,t) \; {\rm e}^{i \bq \cdot \br}$,
where $\cQ$ is the set of wave vectors allowed on different domain topologies. On a torus, $\cQ_{\rm T}$ contains the conventional modes available on a fully periodic domain of size $L\times w$, \textit{i.e.} \hbox{$\cQ_{\rm T}=\{2\pi(m/L,n/w)\}$} with $m=0,1,...,L-1$ and $n=0,1,...,w-1$. The Klein bottle double-cover has toroidal topology over a size of $2L\times w$, \textit{i.e.} geometrically available modes are $\cQ_{\rm K}=\{\pi(m/L,2n/w)\}$ with $m=0,1,...,2L-1$ and $n=0,1,...,w-1$. This set must further be filtered to account for a field-analog version of the topological constraint on pseudoscalars given in Eq.~(\ref{eq:angmod}). Denoting Fourier coefficient by $\tilde{\varphi}(\bq) = \tilde{\varphi}(n,m)$, such a constraint implies~\cite{supp}
\begin{align}\label{eq:pseudoscalar-fourier}
\tilde{\varphi}(m, n) & = - (-1)^{m}\tilde{\varphi}(m,-n).
\end{align}
The mean-field model Eq.~\eqref{eq:meanfield-xy} takes in Fourier space the form of an Ornstein–Uhlenbeck process and reads 
\begin{equation}\label{eq:meanfield_xy_f}
    \partial_t \tilde \varphi(\bq,t) = - \alpha\mathbf{q}^2 \tilde \varphi(\bq,t) + \sqrt{2\beta}\tilde{\xi}(\bq,t)\;,
\end{equation}
where the noise \smash{$\tilde\xi(\bq,t)$} follows from Eq.~\eqref{eq:meanfield-xy} as \smash{$\xi (\br) = \sum_{\bq \in \cQ} \tilde\xi_{\bq}(\br) {\rm e}^{i \bq \cdot \br } $}, and satisfies \smash{$ \mv{ \tilde\xi_{\bq} } =  0$} and \smash{$ \mv{\tilde\xi_{\bq}(t) \tilde\xi_{\bq'}(t')} = \frac{1}{L w} \delta_{\bq+\bq',0} \;\delta (t-t')$}. With these ingredients, fluctuations of the average orientation~$\langle\delta\Phi^2\rangle$ can be computed exactly and read on torus and Klein bottle~\cite{supp} 
\begin{subequations}\label{eq:Theta-meanfield-xy}
    \begin{align}
        \langle{\delta\Phi^2_{\rm T}}\rangle   &= \frac{2\beta }{Lw} t \;,\label{eq:ThetaT_t} \\  
        \langle{\delta\Phi^2_{\rm K}}\rangle  &= \frac{2\beta}{\alpha L^3w}\sum_{\bar{m}=-L/2}^{L/2-1} \frac{1- \exp[-\alpha\,q_{x,\bar{m}}^2t]}{\,q_{x,\bar{m}}^2\sin^2\left(q_{x,\bar{m}}/2\right)} ,\label{eq:ThetaKB_t}
    \end{align}
\end{subequations}
respectively, where $q_{x,\bar{m}} = \pi(2\bar{m}+1)/L$ (assuming even~$L$) and all length scales have been expressed in terms of lattice spacing units~$a$. 
On a torus, Eq.~(\ref{eq:ThetaT_t}) reveals the expected diffusive behavior with a diffusion coefficient that is proportional to the noise amplitude~$\beta$ (Fig.~\ref{fig:fig2}a, inset) and independent of the elastic constant~$\alpha$, confirming the dynamics is dominated by the torque-free soft mode of global spin rotations.


\begin{figure*}[t!]
    \centering
    \begin{tikzpicture}
        \path (0,0) node {\includegraphics[width = 2.05\columnwidth]{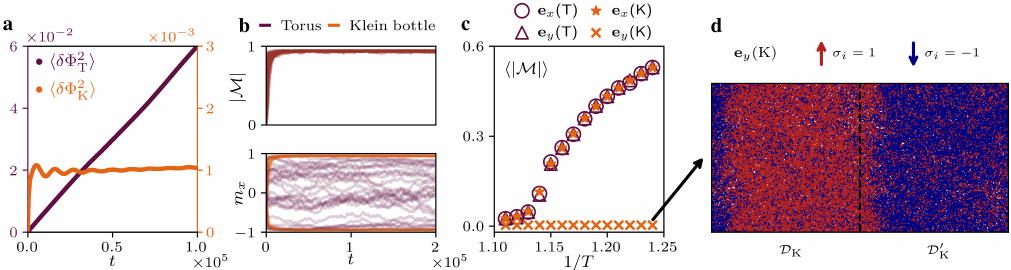}};
    \end{tikzpicture}
    \caption{\textbf{Active particle models on a non-orientable surface.} \textbf{(a)}~Orientation fluctuations in the Vicsek model~\cite{vicsek1995novel} on a torus (purple) and on a Klein bottle (orange). The latter still exhibit a finite bound despite dynamic alignment neighborhoods. \textbf{(b)}~\textit{Top:}~Global ordering in the Vicsek model on torus and Klein bottle.\textit{ Bottom:}~Directional spin polarization diffuses on a torus and exhibits topological pinning behavior on a Klein bottle, analog to the lattice XY model dynamics (Fig.~\ref{fig:fig1}b). \textbf{(c)}~Global orientational order $\langle |\mathcal{M}| \rangle$ in an active Ising model [AIM, see Eq.~(\ref{eq:AIM})] on torus (T) and Klein bottle (K) for Ising spins directed parallel ($\mathbf{e}_x$) and orthogonal ($\mathbf{e}_y$) to the Klein bottle twist axis. The former yields on Klein bottle (orange stars) a similar ordering transition as seen on a torus (empty symbols). Ordering is topologically suppressed in the AIM on a Klein bottle with spin direction $\mathbf{e}_y$ orthogonal to twist axis (orange crosses). \textbf{(d)}~Snapshot of AIM dynamics on Klein bottle with $\mathbf{e}_y$ showing only initialized particles: Topological frustration leads to orientational microphase separation into domains of anti-parallel moving particles and$\langle|\mathcal{M}|\rangle=0$. Parameters: See Methods.
    }
    \label{fig:fig3}
\end{figure*}

Equation~(\ref{eq:ThetaKB_t}) agrees well with the numerical fluctuation dynamics of the XY lattice model on a Klein bottle (Fig.~\ref{fig:fig1}c, black circles) and predicts a scaling collapse that is in excellent agreement with the numerical mean-field results~(Fig.~\ref{fig:fig2}c, black circles). At short time scales ($\alpha\bq^2t \ll 1$), when effects of the local noise have not yet permeated the domain globally, Eq.~(\ref{eq:ThetaKB_t}) yields \hbox{$\langle\delta\Phi^2_{\rm K}\rangle\approx\frac{2\beta}{Lw}t$}~(Fig.~\ref{fig:fig2}c, inset), which gives the same effective diffusion constant as seen with Eq.~(\ref{eq:ThetaT_t}) on a torus for all times. At intermediate times, $t \approx 1/(\alpha\bq^2)$, fluctuations exhibit on a Klein bottle an exponential relaxation that is controlled by the elastic constant $\alpha$ and the system size~$L$. At long times, $\alpha\bq^2t \gg 1$, fluctuations plateau at a constant value 
\begin{equation}\label{eq:ThetaKB_t-long}
\langle\delta \Phi^2_{\rm K}\rangle_{t_\infty}=\frac{2\beta}{\alpha L^3w} \sum_{\bar{m}}\frac{1}{\,q_{x,\bar{m}}^2\sin^2\left(q_{x,\bar{m}}/2\right)}\,, 
\end{equation}
that is controlled by the ratio $\beta/\alpha=D_r/\gamma$ between orientational noise $(D_r)$ and alignment strength~($\gamma$). The latter is a direct consequence of the topologically induced loss of the torque-free diffusive mode and thereby fundamentally distinct from the dynamics on a torus. Even more, this long-time bound on fluctuations still remains in the TDL~\cite{supp},~where
\begin{equation}\label{eq:tdlimitKB}
\langle\delta \Phi^2_{\rm K}\rangle_{t_\infty}\overset{L,w\rightarrow\infty}{\longrightarrow}\frac{\beta p}{6\alpha}=\frac{D_rp}{6\gamma},    
\end{equation}
and  $p = L/w$ is the aspect ratio. The leading correction to this plateau-value is of order $(Lw)^{-1}$, such that numerical results obtained on a domain with $L=w=32$ are already very close to the TDL prediction (Fig.~\ref{fig:fig2}c, black dashed line).

{\it External field analogy and absence of long-range order.} 
The phenomenology observed in Fig.~\ref{fig:fig2}c is reminiscent of the XY model response in the presence of an external field. To investigate this analogy, consider the Hamiltonian~\eqref{eq:domain-energy} for a Klein bottle in the spin-wave limit, and approximate $\mathcal{T}$ as a quadratic confining potential, $\mathcal{T}_{\rmKB}\simeq (A/2) (\varphi_i+\varphi_{k(i)})^2 $, where~$A$ is the potential amplitude and $[i,i(k)]$ are topologically constrained spin pairs on the double-cover. The dynamics of $\varphi_i$ then reads $\partial_t \varphi_{i} = \alpha\Delta\varphi_{i} - A (\varphi_i+\varphi_{k(i)}) + \xi_{i}$ (similar for $\varphi_{k(i)}$). The sum $\phi_i = \varphi_i + \varphi_{k(i)}$ therefore satisfies $\partial_t \phi = \alpha \Delta \phi - 2A \phi +\xi_{\phi}$. The latter is equivalent to the dynamics in an external field $B\equiv A$. Indeed, supplementing the lattice dynamics Eq.~\eqref{eq:xy-dynamics} with an external nematic field interaction $\sim B\sin(2\theta)$ yields in the spin wave limit
\begin{equation}\label{eq:meanfield-xy-field}
    \partial_t \varphi = \alpha \Delta \varphi - 2B \varphi +\sqrt{2\beta} \xi.
\end{equation}
On a torus, the dynamics of fluctuations $\langle \delta \Phi^2 _T\rangle^{B}$ follow from Eq.~\eqref{eq:meanfield-xy-field} as~\cite{gardiner1985handbook} \smash{$\mv{\delta \Phi^2_{\rm T}}^B = \frac{\beta}{BL w} \left[ 1 - {\rm e}^{-2Bt}\right]$}, which reduces to Eq.~\eqref{eq:ThetaT_t} in the limit of $B\to0$ and saturates at long times to a constant value \smash{$\langle \delta\Phi^2_{\rm T}\rangle^B_{t_{\infty}}=\alpha/(B L w)$}. Therefore, an external field on a torus bounds fluctuations in a similar way as the topological constraints imposed by a Klein bottle domain. It is instructive to use Eq.~(\ref{eq:ThetaKB_t-long}) and set \smash{$\langle\delta\Phi^2_{\rm T}\rangle^B_{t_{\infty}}\simeq\langle\delta\Phi^2_{\rm K}\rangle_{t_{\infty}}$}, which yields an effective field strength
\begin{equation}\label{eq:rescaling-B-K}
    \bar{B}=\frac{\alpha}{Lw\langle\delta\Phi^2_{\rm K}\rangle_{t_{\infty}}}.
\end{equation}
that leads on a torus to an equivalent saturation value of fluctuations~(Fig.~\ref{fig:fig2}d, red solid line) as seen on the Klein bottle~(Fig.~\ref{fig:fig2}d, blue solid line). Because \smash{$\langle\delta\Phi^2_{\rm K}\rangle_{t_{\infty}}$} remains finite in the TDL [see Eq.~(\ref{eq:tdlimitKB})], Eq.~(\ref{eq:rescaling-B-K}) implies $\bar B \to 0$ as $L,w\rightarrow\infty$, which suggests the Klein bottle topology does not lead to long-range order in the TDL -- a proposition that we prove explicitly in~\cite{supp}.

{\it Ordering in active models on non-orientable domains.} We finally discuss how our results extend to canonical \textit{non-equilibrium} systems with ferromagnetic interactions. To this end, we replace passive spins with active particles that move in the direction of their spin orientation. As a result, alignment interaction neighborhoods of each particle are constantly reshuffled~\cite{chate2020dry}. While each particle is represented twice on the double-cover $\Omega=\cD\cup\cD'$ -- once as the initialized particle, a second time as an image to enforce topological constraints -- it can now move between double-cover sub-domains $\cD$ and~$\cD'$. During simulations, we keep track of this distinction and compute averages using only initialized particles~(see~Methods). 

We first consider the Vicsek model with continuous spin symmetry~\cite{vicsek1995novel,gregoire2004onset,chate2008collective,ginelli2016physics} in which particles update their orientation in discrete time steps according to \smash{$\theta_i^{t+1} = {\rm arg}\left[ \sum_{j \in \mathcal{N}_i} {\rm e}^{i \theta_j^t} \right] + \eta \xi_i^{t+1}$},
where $\mathcal{N}_i$ contains particle $i$ and its metrical neighbors (see Methods), $\eta$ is the noise amplitude, and $\xi\in[-\pi, \pi]$ is a uniformly distributed random number. Positions $\{\br_i\}$ are updated according to \hbox{$\br_i^{t+1} = \br_i^{t} + d_0 [\cos\theta_i^{t+1},\sin\theta_i^{t+1}]$} for some step length $d_0$. Despite the motion of particles across the Klein bottle surface and dynamic neighborhood changes, orientational order still arises in this model preferentially along the twist-axis direction (Fig.~\ref{fig:fig3}b and Movie~1~\cite{supp}) and average orientation fluctuations of tracked particles remain bounded~(Fig.~\ref{fig:fig3}a), demonstrating the robustness of the topological caging phenomenology. As for the equilibrium case, we are interested in understanding the analogy of the boundedness of fluctuations on the Klein bottle with the impact of an external field. While latter leads on a torus to a plateauing structure factor $S(q)$ at small $q$~\cite{brambati2022signatures}, we find on the Klein bottle numerically a structure factor that remains scale-free~\cite{supp}, suggesting that -- as in the passive case -- non-orientable surface topology does in the TDL not act like an external field. 

As a second example, we consider an active Ising model (AIM)~\cite{solon2013revisiting,solon2015flocking} in which each particle $i$ carries a discrete spin $\sigma_i=\pm 1$ that flips stochastically with rate $W(\sigma_i \to -\sigma_i ) \propto\exp[-\sigma_im_i/(\rho_iT)]$,
where $T$ is an effective temperature, and \smash{$m_i \equiv \sum_{  j \in \mathcal{N}_i} \sigma_j$} and \smash{$\rho_i = \sum_{j \in \mathcal{N}_i} 1$} are the local magnetization and density, respectively. Particles update their position according to
\begin{equation}\label{eq:AIM} 
\dot \br_i  =  v_0 \sigma_i  {\bf e}_{\nu} + \sqrt{2D} \boldsymbol{\zeta}_i \;,
\end{equation}
where $v_0$ is the self-propulsion speed, $D$ is the diffusion coefficient with Gaussian white noise $\boldsymbol{\zeta}_i$, and ${\bf e}_{\nu}$ is a fixed unit vector parallel ($\nu=x$), or orthogonal ($\nu=y$), to the Klein bottle twist axis. The AIM on a Klein bottle with particles moving parallel to the twist-axis (Fig.~\ref{fig:fig3}c, orange stars) exhibits a transition to an orientationally ordered ($\langle|\mathcal{M}|\rangle>0$) polar liquid that is indistinguishable from the known analog transition of this model on a torus~\cite{solon2015phase}~(Fig.~\ref{fig:fig3}c, empty symbols). In contrast,  if particles have to move orthogonal to the twist axis (Eq.~(\ref{eq:AIM}) with~$\nu=y$), the ordering transition is fully suppressed by the topology of the Klein bottle surface~(Fig.~\ref{fig:fig3}c, orange crosses). To rationalize this, we recall that vector components orthogonal to the twist-axis transform like pseudoscalars on the double-cover of a Klein bottle and can therefore not take a globally constant non-zero value. Hence, a globally ordered polar liquid state that is oriented orthogonal to the twist-axis is topologically suppressed. At all temperatures for which a polar liquid state with $\langle|\mathcal{M}|\rangle>0$ emerges on the torus, the corresponding frustration on a Klein bottle gets resolved by an orientational microphase separation in which particles form two coexisting domains of anti-parallel moving particles. which leads to $\langle|\mathcal{M}|\rangle\approx0$~(Fig.~\ref{fig:fig3}d, and Movie~2~\cite{supp}).

We have shown how the removal of global rotational soft modes by the non-orientability of a surface reshapes the collective statistical properties of systems with orientational degrees of freedom. Using the Klein bottle as a paradigmatic example, we considered an orientable double-cover to reformulate topological constraints as non-local relations, which overcomes the need to treat curved surfaces~\cite{vann21,nebel23} and enabled exact analytic treatments of stochastic models. In a passive XY model, non-orientability gives rise to \textit{topological caging}: Homogeneous ordered states get pinned to the twist axis, which converts toroidal Goldstone mode diffusion into bounded fluctuations. We have shown this bound is controlled by a competition between rotational diffusion and alignment strength that persists in the~TDL. In active models, topological caging remains robust despite a dynamic reshuffling of interaction neighborhoods. In the AIM, non-orientability acts selectively on accessible ordered phases: Motion parallel to the twist axis supports the usual polar liquid transition, whereas spin orientation orthogonal to the twist axis leads to topological frustration and globally disordered microphase separation. Our approach provides a general route for studying stochastic fields and many-body dynamics on non-orientable manifolds and could be applied to generic $p$ ordering systems~\cite{solon2022susceptibility,mietke2022anyonic}, active polar and nematic hydrodynamic models~\cite{patelli2019understanding,chate2020dry,mahault2021long}, or systems in which polar order emerges from non-polar interactions~\cite{dinelli2023non,pisegna2024emergent}. Exploring the impact of non-orientability in these systems may reveal a wider class of collective phenomena in which domain topology can be used to select emergent~phases.

\section*{Acknowledgments}
We thank Siddharth Parameswaran, Ioannis Hadjifrangiskou, Sarah M. Loos, Simone Sotgiu, and Giuseppe Fava for helpful discussions. GS acknowledges support from the ERC Advanced Grant ActBio (funded as UKRI Frontier Research Grant EP/Y033981/1).

\bibliographystyle{apsrev4-2}
\bibliography{biblio}
\ \\
\section*{Methods}
\noindent

{\it Numerical simulations.} We have simulated all models using an Euler scheme with a fixed time step $dt$ on fully periodic domains ${\Omega} = \cD \cup \cD'$ of size $2L\times w$. Averages of the XY dynamics (Figs.~\ref{fig:fig1},\ref{fig:fig2}) are calculated w.l.o.g. over the lattice sites on the domain $\cD$. In simulations of active models (Fig.~\ref{fig:fig3}), we initialize all physical particles on~$\cD$ (and their topological images on $\cD'$). Averages are calculated over the initialized physical particles, while alignment neighborhoods take into account nearby physical particles and topological images.\\

\begin{figure*}[t!]
    \centering
    \includegraphics[width = 2.05\columnwidth]{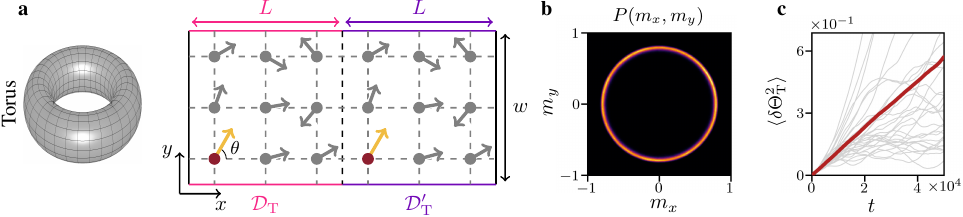}\vspace{-0.2cm}
    \caption{
    \textbf{Fluctuation measurements of the XY model on a torus.} \textbf{(a)}~Trivial double-cover $\Omega_T=\cD_T\cup\cD'_T$ of domain $L\times w$ with toroidal symmetry: All fields satisfy ${\bff}^{\rmT}(x+L, y+w) = \bff^{\rmT}(x,y)$. \textbf{(b)}~Enlarged inset of Fig.~\ref{fig:fig1}b:~Polarization distribution is isotropic due to diffusing Goldstone mode of global homogeneous spin rotations~\cite{goldenfeld2018lectures}. \textbf{(c)}~Enlarged inset of Fig.~\ref{fig:fig1}c:~Fluctuations of average orientations exhibit conventional diffusive dynamics (gray: individual realizations, red: average). Simulation parameters: See Methods.
    }
    \label{fig:fig4}
\end{figure*}

{\it Simulation parameters.}
\begin{itemize}[leftmargin=*]\setlength{\itemsep}{0.1cm}
\item Fig.~\ref{fig:fig1}: Simulations are initialized with random initial conditions. $L=w=128$, $\gamma=1$, $D_r=0.1$, $dt=0.01$, total simulation time $T_{\rm sim}=10^{5}D_r^{-1}$ (panels b,c). $\langle\delta\Theta^2\rangle$-data in panel c is computed by keeping track of the winding number of $\Theta$. The averages are taken over $N_s=30$ independent realizations.
\item Fig.~\ref{fig:fig2}a-c: $L=w=32$, $dt=0.01$. Total simulation time $T_{\rm sim}=10^6$, starting from ordered state \hbox{$\theta(x,y) = 0$}. Averages are computed from $N_s=30$ independent realizations.
\item Fig.~\ref{fig:fig2}d: $\alpha=0.01$, $\beta = 0.005$ ($\langle\Phi^2_K\rangle$) and, \hbox{$\bar B = 5.8\cdot 10^{-5}$} as determined from Eq.~\eqref{eq:rescaling-B-K}, $\beta = 0.005$ ($\langle\Phi^2_T\rangle^{\bar{B}}$). Averages are taken over $N_s=30$ independent realizations.
\item Fig.~\ref{fig:fig3}a-b:~We initialize $N=\rho_0 Lw$ particles with random positions and orientations on $\cD$, and then generate their double-cover images on $\cD'$. Simulation parameters: $L=w=256$, $\rho_0= 1.0$, $d_0=0.5$, $\eta=0.1$. The set $\mathcal{N}_i$ of particles interacting with particle $i$ is defined as $\mathcal{N}_i = \{j \;{\rm s.t.}\; |\br_j -\br_i| < 1\;, j\in \Omega = \cD_K\cup \cD_K'  \}$. Averages are taken over $N_s=30$ independent realizations.
\item Fig.~\ref{fig:fig3}c-d:~Particle initialization and interaction neighborhoods as in Fig~\ref{fig:fig3}a. $L_x=L_y=200$, $\rho_0=1.2$, $v_0=0.5$, $dt = 1/{\rm exp}(\beta)$, and $D=5$. Panel d corresponds to $\beta=1.24$. Note that here $\beta$ represents the inverse temperature, $i.e.$ $\beta=1/T$. Spin flips stochastically with rate $W(\sigma_i \to -\sigma_i ) = \Gamma \exp[-\sigma_im_i/(\rho_iT)]$, with $\Gamma=0.5$.
\item Fig.~\ref{fig:fig4}b,c. Same simulation parameters and analysis methods as used in Fig.~\ref{fig:fig1}.
\end{itemize}
\end{document}


\title{\Large Orientational order on non-orientable domains \\ \large Supplementary Materials}

\author{Gianmarco Spera}
\affiliation{Rudolf Peierls Centre for Theoretical Physics, University of Oxford, Oxford OX1 3PU, United Kingdom}
\author{Axel Fotso Ndefo}
\affiliation{Rudolf Peierls Centre for Theoretical Physics, University of Oxford, Oxford OX1 3PU, United Kingdom}
\author{Keaton J. Burns}
\affiliation{Department of Mathematics, Massachusetts Institute of Technology, Cambridge, MA 02139}
\affiliation{Center for Computational Astrophysics, Flatiron Institute, New York, NY 10010}
\author{Alexander Mietke}
\email{alexander.mietke@physics.ox.ac.uk}
\affiliation{Rudolf Peierls Centre for Theoretical Physics, University of Oxford, Oxford OX1 3PU, United Kingdom}
\date{\today}

\maketitle
\tableofcontents

\onecolumngrid
\appendix
\setcounter{figure}{0}
\renewcommand{\thefigure}{S\arabic{figure}}
\setcounter{table}{0}
\renewcommand{\thetable}{S\arabic{table}}

\vspace{\fill}

The Supplemental Material contains detailed derivations of analytic results and additional information about numerical results presented in the main text. In Sec.~\ref{app:meanfield-xy}, we derive the key analytic results presented in the main text. In Sec.~\ref{sec:order}, we show that the spin-wave Hamiltonian on Klein bottle remains Gaussian and derive an explicit expression for the spatial spin correlation function. In Sec.~\ref{sec:structure}, we report numerical measurements of the structure factor in the active Vicsek model on a Klein bottle and discuss implications for the analogy between non-orientable domain topology and external fields.

\newpage

\section{Stochastic mean-field theory}\label{app:meanfield-xy}
In this section, we derive Eqs.~(\ref{eq:ThetaKB_t}) from the main text -- an exact expression for the time-dependent fluctuations of average global orientation in a mean-field XY model on a Klein-bottle. Recalling Eq.~(\ref{eq:meanfield-xy}) from the main text, the stochastic dynamics of local orientations $\varphi(\br,t)$ in a spin-wave approximation reads
\begin{equation}\label{app-eq:mean-field-dynamics}
    \partial_t \varphi(\br, t) = \alpha \Delta \varphi(\br,t) + \sqrt{2 \beta} \xi(\br,t)\;,
\end{equation}
where $\xi$ is a centered Gaussian white noise with $\langle \xi(\br,t) \xi(\br',t')\rangle = \delta(t-t') \delta(\br -\br')$. Average orientations $\Phi$ are defined as 
\begin{equation}\label{app-eq:phi-tot}
    \Phi(t) = \frac{1}{Lw} \int \rmd \br \; \varphi(\br) =  \frac{a^2}{Lw} \sum_{\br} \varphi(\br)\;,
\end{equation}
where $L\times w$ is the system size, $a$ is the lattice spacing, and the sum runs over lattice sites. For brevity, we express in the following all lengths in terms of $a$. To characterize the dynamics of~$\Phi(t)$, we want to compute its fluctuations defined by
\begin{equation}\label{app-eq:msd}
    \langle\delta\Phi^2 \rangle := \langle [\Phi(t+t_0) - \Phi(t_0)]^2\rangle_{t_0}\;,
\end{equation}
where the average is taken over different initial times $t_0$. To determine Eq.~\eqref{app-eq:msd} from the dynamics Eq.~\eqref{app-eq:mean-field-dynamics}, we go to Fourier space, where we define the discrete Fourier transformation $\tilde f(\bq)$ of a function $f(\br)$, as well as the inverse transformation, on a domain of size $L\times w$ as
\begin{equation}
    \tilde f(\bq, t) = \frac{1}{Lw} \sum_{\br} f(\br, t) {\rm e}^{-i \bq \cdot \br}\quad\Leftrightarrow\quad f(\br, t) =  \sum_{\bq} \tilde f(\bq, t) {\rm e}^{i \bq \cdot \br}.
\end{equation}
Straightforward algebra leads to Eq.~(\ref{eq:meanfield_xy_f}) in the main text, where the Fourier-transformed noise satisfies
\begin{equation}\label{app-eq:mf-xy-fourier}
     \mv{\tilde \xi{(\bq,t)} \tilde \xi{(\bq',t)'}} = \frac{1}{L w} \delta_{\bq+\bq',0} \;\delta (t-t').
\end{equation}

\subsection{Fourier mode symmetries of scalar and pseudoscalar fields on a Klein bottle}
As stated with Eq.~(\ref{eq:pseudoscalar-fourier}) in the main text, the non-local constraints imposed by a Klein bottle topology onto an orientable double-cover imply simple conditions on Fourier coefficients. To derive those, recall first that the set of geometrically allowed modes on the Klein bottle double-cover of size $2L\times w$ is
\begin{equation}\label{eq:QK}
\mathcal{Q}_{\rm K}=\{\mathbf{q}=\pi(m\mathbf{e}_x/L+2n\mathbf{e}_y/w)|m,n\in\mathbb{Z}\}\mod 2\pi,
\end{equation}
which takes into account that fields have only $2L$-periodicity along the twist axis (chosen w.l.o.g. as the $x$-axis on our work). The set given in Eq.~(\ref{eq:QK}) is subject to additional constraints implied by the Klein bottle topology:
\begin{itemize}
\item \textit{Scalar fields:} On the Klein bottle double-cover, a scalar field $\varphi$ satisfies $\varphi(x, y) = \varphi(x+L, w-y)$. To impose this constraint, consider
\begin{align}      
\varphi(x, y) - \varphi(x+L, w-y) & = \sum_{\bq} \tilde \varphi(\bq) \left[ {\rm e}^{i (q_x x + q_y y)} - {\rm e}^{i [q_x (x+L) + q_y (w-y)]} \right]  \notag\\
& = \sum_{\bq} \tilde \varphi(\bq) \left[ {\rm e}^{i (q_{x} x + q_y y)} - (-1)^m{\rm e}^{i(q_x x - q_y y)} \right] \;,\label{app-eq:scalar}  
\end{align}
where we have used ${\rm e}^{i q_y w} = 1$ and  ${\rm e}^{i q_x L} = (-1)^m$. Equation~\eqref{app-eq:scalar} vanishes only if $\tilde \varphi(m, n) = (-1)^{m} \tilde \varphi(m,-n)$, where we write for convenience $\tilde \varphi(\bq)\equiv\tilde\varphi(m, n)$ and mode indices ($m,n$) map to mode vectors $\bq$ as described in Eq.~(\ref{eq:QK}).
\item \textit{Pseudoscalar fields:} On the other hand, pseudoscalar fields on a Klein bottle satisfy $\varphi(x, y) = - \varphi(x+L, w-y)$ (see Eq.~(\ref{eq:angmod}) in the main text). Consider therefore
\begin{align}       
\varphi(x, y) + \varphi(x+L, w-y) & = \sum_{\bq} \tilde \varphi(\bq) \left[ {\rm e}^{i (q_x x + q_y y)} + {\rm e}^{i [q_x (x+L) + q_y (w-y)]} \right]\notag\\
& = \sum_{\bq} \tilde \varphi(\bq) \left[ {\rm e}^{i (q_x x + q_y y)} + (-1)^m{\rm e}^{i [q_x x - q_y y]} \right],\label{app-eq:pseudoscalar}
\end{align}
which only vanishes if the Fourier modes satisfy $\tilde \varphi(m, n) = -(-1)^{m} \tilde \varphi(m,-n)$, corresponding to the identify Eq.~(\ref{eq:pseudoscalar-fourier}) stated in the main text.
\end{itemize}

We therefore find that Fourier coefficients of a field $\varphi(\mathbf{r})$ on a Klein bottle double-cover must satisfy
\begin{subequations}
    \begin{align}        
    \tilde \varphi(m, n) & = (-1)^{m} \tilde \varphi(m,-n) &&\hspace{-3cm} {\rm if}\; \varphi \; \text{is a scalar}\;, \label{app-eq:scalar-fourier} \\
    \tilde \varphi(m, n) & = - (-1)^{m} \tilde \varphi(m,-n) &&\hspace{-3cm} {\rm if}\; \varphi \; \text{is a pseudoscalar}\;. \label{app-eq:pseudoscalar-fourier}
    \end{align}
\end{subequations}
The orientation fields discussed in the main text are pseudoscalars, for which Eq.~(\ref{app-eq:pseudoscalar-fourier}) implies modes that are even (odd) along the twist axis (related to $m$) have an odd (even) symmetry along the $y$-axis (related to $n$).

\subsection{Orientation fluctuations on a Torus}
For reference, we first re-derive global orientation fluctuations $\langle \delta \Phi^2 _{\rmT}\rangle$ of the XY model on a torus. The computation of is straightforward by noticing that $\varphi_{\bq=0}$ is related to the global magnetization defined in Eq.~\eqref{app-eq:phi-tot}:
\begin{equation}
    \varphi_{\bq = 0}(t) = \left. \frac{1}{Lw} \sum_{\br \in \cD} \varphi (\br) {\rm e}^{i \bq \textbf{r}} \right |_{\bq = 0} = \frac{1}{Lw} \sum_{\br \in \cD} \varphi (\br) = \Phi(t)\;,
\end{equation}
where the normalization $1/{Lw}$ accounts for the number of modes in the inverse Fourier transform. Therefore, the time evolution of the global average orientation $\Phi(t)$ is described by Eq.~\eqref{app-eq:mf-xy-fourier} for $\bq=0$, which reads
\begin{equation}
    \partial_t \Phi(t) = \sqrt{2\beta} \xi_{\bq =0}(t)\;.
\end{equation}
The latter is simply the equation of a Brownian motion and thus satisfies 
\begin{equation}\label{app-eq:msd-xy-torus}
    \mv{ \Phi(t)_{\rmT}} = 0 \;, \quad \text{and}\quad \langle \delta \Phi^2 _{\rmT}\rangle = \frac{2\beta}{Lw} t\;,
\end{equation}
which is Eq.~(\ref{eq:ThetaT_t}) of the main text.

\subsection{Orientation fluctuations on a Klein bottle}\label{app:meanfield-xy-kb}
We now consider the XY model on a Klein bottle, whose topology imposes the mode constraints Eq.~\eqref{app-eq:pseudoscalar-fourier} on a pseudoscalar orientation field $\varphi (\br)$. With these constraints, the Fourier expansion of $\varphi$ becomes
\begin{equation}\label{app-eq:fourier-kb}
    \varphi(\br) = \sum_{\bq \in \mathcal{Q}_{\rm K}} \Tilde{\varphi}(\bq) {\rm e}^{i q_x x} t_{q_x}(q_y y)\;,
\end{equation} 
where the function $t_{q_x}$ is defined as 
\begin{equation}
    t_{q_x}(q_y y ) \equiv \begin{cases}
        \sin(q_y y) \quad\text{if $m$ is even}  \\ 
        \cos(q_y y) \quad\text{if $m$ is odd}
    \end{cases}\;,
\end{equation}
and $\mathcal{Q}_{\rm K}$ is the set of geometrically allowed modes given in Eq~(\ref{eq:QK}). Using the mode-constrained expansion Eq.~(\ref{app-eq:fourier-kb}), the average orientation $\Phi$ given in Eq.~(\ref{app-eq:phi-tot}) becomes
\begin{equation}\label{app-eq:phi-kb}
    \Phi = \frac{1}{Lw} \sum_{x=0}^{L-1} \sum_{y=0}^{w-1}  \sum_{\bq \in \mathcal{Q}_{\rm K}} \Tilde{\varphi}(\bq) {\rm e}^{i q_x x} t_{q_x}(q_y y) \;.
\end{equation}
We simplify Eq.~\eqref{app-eq:phi-kb} by performing the sum over $x$ and $y$. The sum over $x$ reads 
\begin{equation}\label{app-eq:sum-x1}
    \sum_{x=0}^{L-1} {\rm e}^{i q_{x,m} x} = \sum_{x =0}^{L-1} {\rm e}^{i \frac{m\pi}{L} x} = 
    L \delta_{m,0} +  (1 - \delta_{m,0}) \sum_{x=0}^{L-1} {\rm e}^{i \frac{m\pi}{L} x} 
    = L \delta_{m,0} + \frac{i(-1)^{m+1}{\rm e}^{-i\frac{m\pi}{2L}}}{\sin(\frac{m\pi}{2L}) } (1 - \delta_{m,0})\;,
\end{equation} 
where we used the geometric series identity $\sum_{n=0}^{N-1} x^n = (1 - x^{N})/(1-x)$ and separated its contribution for $m=0$ and $m\neq0$. Computing the sum over $y$ in Eq.~\eqref{app-eq:phi-kb} requires a case distinction. When $m$ is even, the sum reads 
\begin{equation}\label{app-eq:sum-sin-even}
    \sum_{y=0}^{w-1} t_{q_x}(q_y y) = \sum_{y=0}^{w-1} \sin(q_y y) = \frac{1}{2i}\sum_{y=0}^{w-1} \left({\rm e}^{iq_y y} - {\rm e}^{-iq_y y}\right) = \frac{1}{2i} \left[ \frac{ 1- (-1)^{2n\pi} }{1 - {\rm e}^{i \frac{2n\pi}{w} } } - \frac{1-(-1)^{-2n\pi}}{1 - {\rm e}^{-i \frac{2n\pi}{w} }} \right] = 0\;. 
\end{equation}
Thus, all contributions in Eq.~\eqref{app-eq:phi-kb} with even $m$ vanish. On the other hand, when $m$ is odd,  the sum over $y$ reads
\begin{equation}\label{app-eq:sum-cos-odd}
    \begin{split}        
    \sum_y t_{q_x}(q_y y)  = \sum_{y=0}^{w-1} \cos(q_y y) 
    = w \delta_{n,0} + \frac{1}{2} \left[ \frac{ 1 - (-1)^{2n\pi} }{1 - {\rm e}^{i \frac{i2n\pi}{w} } } + \frac{1 - (-1)^{-2n\pi}}{1 - {\rm e}^{-i \frac{i2n\pi}{w} }} \right] (1 - \delta_{n,0})  = w \delta_{n,0} \;.
    \end{split}    
\end{equation}
This implies the only non-vanishing contribution from the sum over $y$ comes from $n=0$. Using Eqs.~\eqref{app-eq:sum-x1}--\eqref{app-eq:sum-cos-odd}, Eq.~\eqref{app-eq:phi-kb} becomes
\begin{equation}\label{app-eq:phi-fourier}
    \Phi = 
    \frac{1}{L} \sum_{\substack{m\; \text{odd} \\n=0 } } \tilde \varphi(q_x, 0) \left[ \frac{i{\rm e}^{-i\frac{m\pi}{2L}}}{\sin(\frac{m\pi}{2L}) }\right].
\end{equation}
Finally, $\langle \delta \Phi^2_{\rm K}\rangle$ follows from Eq.~\eqref{app-eq:phi-fourier} as 
\begin{equation}\label{app-eq:phi-fourier-2}
    \begin{split}        
    \langle \delta \Phi^2_{\rm K} \rangle & = \langle [\Phi(t) - \Phi(0)]^2 \rangle = \langle [\Phi(t) - \Phi(0)][\Phi(t) - \Phi(0)]^* \rangle  \\ & 
    = \frac{1}{L^2}   \sum_{m\; \text{odd}} \sum_{m'\; \text{odd}} \langle [\tilde \varphi(q_x, 0, t) - \tilde \varphi(q_x, 0, 0) ]  [\tilde\varphi^*(q_x', 0,t) - \tilde\varphi^*(q_x', 0, 0)] \rangle \left[ \frac{i{\rm e}^{-i\frac{m\pi}{2L}}}{\sin(\frac{m\pi}{2L}) }\right]\left[ \frac{-i{\rm e}^{i\frac{m'\pi}{2L}}}{\sin(\frac{m'\pi}{2L}) }\right] \;, 
    \end{split}
\end{equation}
where $\tilde \varphi^*$ is the complex conjugate of $\tilde \varphi$. As $\tilde \varphi$ obeys the Ornstein-Uhlenbeck dynamics Eq.~\eqref{app-eq:mf-xy-fourier}, its variance satisfies
\begin{equation}\label{app-eq:variance-ou}
    \langle  [\tilde\varphi(q_{x}, 0, t) - \tilde \varphi(q_x, 0, 0) ]  [\tilde\varphi^*(q_x', 0,t) - \tilde \varphi^*(q_x', 0, 0)] \rangle = \frac{2\beta }{L w} \frac{1 - {\rm e}^{ - \alpha q^2t}}{\alpha q^2} {\delta_{m,m'}}\;,
\end{equation}
where $q= |\bq|$. Substituting Eq.~\eqref{app-eq:variance-ou} into Eq.~\eqref{app-eq:phi-fourier-2}, we obtain
\begin{equation}\label{app-eq:diff-kb}
    \langle {\delta\Phi^2}_{\rm K} \rangle  =  \frac{2\beta}{L^3w} \sum_{\substack{m\; \text{odd} \\n=0 } }\frac{1- {\rm e}^{- \alpha q^2t}}{\alpha q^2}  {\left[ \sin\left(\frac{m\pi}{2L}\right)\right]^{-2}}.
\end{equation}
Noting that $\bq=\frac{\pi m}{L}\mathbf{e}_x$ when $n=0$, we see that Eq.~(\ref{app-eq:diff-kb}) corresponds to Eq.~(\ref{eq:ThetaKB_t}) stated in the main text. 

\subsection{Boundedness of fluctuations in the thermodynamic limit}
We now show that Eq.~\eqref{app-eq:diff-kb} plateaus at a finite value at long times and in the thermodynamics limit, $i.e.$ fluctuations remain always bounded. For convenience, we wrote the long term limit of Eq.~\eqref{app-eq:diff-kb} in the main text [see Eq.~(\ref{eq:ThetaKB_t-long})] as
\begin{equation}\label{app-eq:Theta-meanfield-xy-kb-long}
\ \hspace{-0.1cm}\langle \delta \Phi^2_{\rm K}\rangle\rightarrow\langle\delta \Phi^2_{\rm K}\rangle_{t_\infty}=\frac{2\alpha}{\beta L^3w} \sum_{\bar{m}\in\mathbb{Z}}\frac{1}{\,q_{x,\bar{m}}^2\sin^2\left(q_{x,\bar{m}}/2\right)} \;, 
\end{equation}
where $q_{x,\bar{m}}=\pi(2\bar{m}+1)/L$. For finite system sizes, only a finite number of modes contribute to the sum in Eq.~\eqref{app-eq:Theta-meanfield-xy-kb-long}, which trivially implies that fluctuations $\langle\delta \Phi^2_{\rm K}\rangle_{t_\infty}$ plateau at a finite value. To determine what happens in the the thermodynamic limit ($L,w \to \infty$ with $p=L/w$ constant), we expand $\sin^2(q_{x,\bar{m}}/2)$ to leading order in system size. Equation~\eqref{app-eq:Theta-meanfield-xy-kb-long} then becomes
\begin{equation}\label{app-eq:sum-kb-intermediate}
    \langle\delta \Phi^2_{\rm K}\rangle_{t_\infty}=\frac{2\alpha}{\beta L^3w} \sum_{\bar{m}\in\mathbb{Z}}\frac{1}{\,q_{x,\bar{m}}^2\sin^2\left(q_{x,\bar{m}}/2\right)} \simeq \frac{2\alpha}{\beta L^3 w } \sum_{\bar{m} \in \mathbb{Z}} \left[ \frac{4}{q^4_{x,\bar{m}}} + \mathcal{O}(q_{x,m}^2) \right] = \frac{8\alpha L}{\pi^4 w}\sum_{\bar{m} \in \mathbb{Z}} \frac{1}{(2\bar{m}+1)^4} + \mathcal{O}\left(\frac{1}{Lw}\right)\;.
\end{equation}
In the thermodynamic limit, the sum in Eq.~\eqref{app-eq:sum-kb-intermediate} is 
\begin{equation}
    \sum_{\bar{m}\in\mathbb{Z}} \frac{1}{(2\bar{m}+1)^4} = 2\sum_{\bar{m} =0}^{\infty} \frac{1}{(2\bar{m}+1)^4} = \frac{\pi^4}{48}\,
\end{equation}
such that Eq.\eqref{app-eq:sum-kb-intermediate} becomes in the thermodynamic limit
\begin{equation}\label{app-eq:tdlimitKB}
    \lim_{ \substack{L \to \infty \\ L/w = p } } \langle\delta \Phi^2_{\rm K}\rangle_{t_\infty} = \frac{\beta p}{6 \alpha} + \mathcal{O}\left( \frac{1}{Lw}\right)\;,
\end{equation}
which is Eq.~(\ref{eq:tdlimitKB}) of the main text. 

\section{Orientational ordering on the Klein bottle in the thermodynamic limit}\label{sec:order}
In this section, we derive the Hamiltonian that corresponds to the spin-wave approximation of the XY model on a Klein bottle. First, we show that the Hamiltonian remains Gaussian, and then that spatian correlations still decay as a power law, \textit{i.e.} the Klein bottle topology does \textit{not} lead to true long-range order in the thermodynamic limit. 

\subsection{Spin-wave Hamiltonian on a Klein bottle}\label{sec:hamiltonian}
In Sec.~\ref{app:meanfield-xy-kb}, we have derived a Fourier expansion of a pseudoscalar field $\varphi(x,y)$,
\begin{equation}\label{app-eq:theta-fourier}
    \varphi(x,y) = \sum_{\substack{m\; \text{odd} \\n} }\tilde\varphi(m,n) {\rm e}^{i q_x x} \cos(q_y y) + \sum_{\substack{m\; \text{even} \\n } }\tilde\varphi(m,n) {\rm e}^{i q_x x} \sin(q_y y)\;, \quad\text{with}\quad q_x = \frac{m \pi}{L},\ q_y = \frac{2n\pi}{w}\;,
\end{equation}
which satisfies by construction the topological constraints of a Klein bottle surface. The expansion Eq.~(\ref{app:meanfield-xy-kb}) will be used in the following to understand how the topological mode constraints affect the Hamiltonian associated with the spin-wave approximation. The latter reads
\begin{equation}\label{app-eq:hamiltonian}
    \mathcal{H} = \frac{\alpha}{2} \int {\rm d} \br(\nabla\varphi)^2 = \frac{\alpha}{2} \sum_{x=0}^{L-1} \sum_{y=0}^{w-1} \left[ \left( \partial_x\varphi\right)^2 + \left( \partial_y\varphi\right)^2 \right] \;,
\end{equation}
where $\alpha$ is the elastic constant the alignment interactions and the sum runs over the available lattice sites. To express this Hamiltonian on a Klein bottle, we use Eq.~(\ref{app-eq:theta-fourier}) to evaluate the terms $(\partial_x\varphi)^2$ and $(\partial_y\varphi)^2$. Specifically, we have
\begin{equation}
    \begin{split}
            (\partial_x \varphi )^2  = & \sum_{\substack{m,m'\;{\rm odd} \\ n,n'}}\tilde\varphi(m,n)\tilde\varphi(m',n') (-q_x q_x') {\rm e}^{i(q_x+q_x')x} \cos(q_y y) \cos(q_y' y)  \\
            + & \sum_{\substack{m,m'\;{\rm even} \\ n,n'}}\tilde\varphi(m,n)\tilde\varphi(m',n') (-q_x q_x') {\rm e}^{i(q_x+q_x')x} \sin(q_y y) \sin(q_y' y) \\
            + & \sum_{\substack{m\;{\rm odd}\\m'\;{\rm even} \\ n,n'}}\tilde\varphi(m,n)\tilde\varphi(m',n') (-q_x q_x') {\rm e}^{i(q_x+q_x')x} \cos(q_y y) \sin(q_y' y) \\
            + & \sum_{\substack{m\;{\rm even}\\m'\;{\rm odd} \\ n,n'}}\tilde\varphi(m,n)\tilde\varphi(m',n') (-q_x q_x') {\rm e}^{i(q_x+q_x')x} \sin(q_y y) \cos(q_y' y) \;,
    \end{split}
\end{equation}
which we re-write, using standard trigonometric identities, as
\begin{equation}\label{eq:dpx_sq}
    \begin{split}
            (\partial_x \varphi )^2  =&  \sum_{\substack{m,m'\;{\rm odd} \\ n,n'}}\tilde\varphi(m,n)\tilde\varphi(m',n') (-q_x q_x') {\rm e}^{i(q_x+q_x')x} \frac{1}{2} \left\{ \cos\left[(q_y-q_y')y\right] + \cos\left[(q_y+q_y')y\right] \right\}   \\
            + & \sum_{\substack{m,m'\;{\rm even} \\ n,n'}}\tilde\varphi(m,n)\tilde\varphi(m',n') (-q_x q_x') {\rm e}^{i(q_x+q_x')x} \frac{1}{2} \left\{ \cos\left[(q_y-q_y')y\right] - \cos\left[(q_y+q_y')y\right] \right\} \\
            + & \sum_{\substack{m\;{\rm odd}\\m'\;{\rm even} \\ n,n'}}\tilde\varphi(m,n)\tilde\varphi(m',n') (-q_x q_x') {\rm e}^{i(q_x+q_x')x} \frac{1}{2} \left\{ \sin\left[(q_y+q_y')y\right] - \sin\left[(q_y-q_y')y\right] \right\} \\
            + & \sum_{\substack{m\;{\rm even}\\m'\;{\rm odd} \\ n,n'}}\tilde\varphi(m,n)\tilde\varphi(m',n') (-q_x q_x') {\rm e}^{i(q_x+q_x')x} \frac{1}{2} \left\{\sin\left[(q_y+q_y')y\right] + \sin\left[(q_y-q_y')y\right] \right\}\;.
    \end{split}
\end{equation}
Similarly, we obtain 
\begin{equation}\label{eq:dpy_sq}
    \begin{split}
            (\partial_y \varphi )^2  = & \sum_{\substack{m,m'\;{\rm odd} \\ n,n'}}\tilde\varphi(m,n)\tilde\varphi(m',n') (q_y q_y') {\rm e}^{i(q_x+q_x')x} \frac{1}{2} \left\{ \cos\left[(q_y-q_y')y\right] - \cos\left[(q_y+q_y')y\right] \right\}     \\
            + & \sum_{\substack{m,m'\;{\rm even} \\ n,n'}}\tilde\varphi(m,n)\tilde\varphi(m',n') (q_y q_y') {\rm e}^{i(q_x+q_x')x} \frac{1}{2} \left\{ \cos\left[(q_y-q_y')y\right] + \cos\left[(q_y+q_y')y\right] \right\} \\
            + & \sum_{\substack{m\;{\rm odd}\\m'\;{\rm even} \\ n,n'}}\tilde\varphi(m,n)\tilde\varphi(m',n') (-q_y q_y') {\rm e}^{i(q_x+q_x')x} \frac{1}{2} \left\{ \sin\left[(q_y+q_y')y\right] - \sin\left[(q_y-q_y')y\right] \right\} \\
            + & \sum_{\substack{m\;{\rm even}\\m'\;{\rm odd} \\ n,n'}}\tilde\varphi(m,n)\tilde\varphi(m',n') (-q_y  q_y') {\rm e}^{i(q_x+q_x')x} \frac{1}{2} \left\{\sin\left[(q_y+q_y')y\right] + \sin\left[(q_y-q_y')y\right] \right\}.\\
    \end{split}
\end{equation}
To obtain the full Hamiltonian Eq.~(\ref{app-eq:hamiltonian}) in Fourier space, we next compute the sum over $x$ and $y$. Summations over $x$ in Eqs.~(\ref{eq:dpx_sq}) and (\ref{eq:dpy_sq}) contribute
\begin{equation}\label{app-eq:sum-x}
    \sum_{x=0}^{L-1} {\rm e}^{i(q_x+q_x')x} = \sum_{x=0}^{L-1} {\rm e}^{i\frac{\pi}{L}(m+m')x} = \frac{1- {\rm e}^{i \pi (m+m')}}{1- {\rm e}^{i \pi (m+m')/L}} = L \delta_{m,-m'} + ( 1- \delta_{m,-m'}) g(m,m')
\end{equation}
where the function $g(m,m')$ is defined as
\begin{equation}
    g(m,m') = \begin{cases}
        0 \hspace{2.15cm} \text{if } m+m'\text{ is even}  \\ 
        \frac{2}{1-{\rm e}^{i \pi (m+m')/L}}\hspace{0.3cm}  \text{if } m+m'\text{ is odd}
    \end{cases}.
\end{equation}
Summations over $y$ in Eqs.~(\ref{eq:dpx_sq}) and (\ref{eq:dpy_sq}) contribute
\begin{equation}\label{app-eq:cos-all}
\sum_{y=0}^{w - 1}\cos\left[(q_y\pm q_y')y\right]=\sum_{y=0}^{w-1} \cos\left[\frac{2\pi}{w} (n\pm n')y\right] = w \delta_{n,\mp n'},
\end{equation}
and
\begin{equation}\label{app-eq:sin-sum}
\sum_{y=0}^{w - 1}\sin\left[(q_y\pm q_y')y\right]=\sum_{y=0}^{w - 1} \sin\left[\frac{2\pi}{w}(n\pm n')y\right] = 0 \;.
\end{equation}
First, we note that all the vanishing sums given in Eq.~\eqref{app-eq:sin-sum} correspond in Eqs.~(\ref{eq:dpx_sq}) and (\ref{eq:dpy_sq}) to summations in which $m$ and~$m'$ have opposite parity. We are thus only left with contributions in which $m$ and $m'$ are both odd or even. The sum over odd modes $m$ and $m'$ can be rewritten as 
\begin{equation}\label{app-eq:ham-odd}
    \sum_{x=0}^{L-1}\sum_{y=0}^{w-1} \left[ \left(\partial_x \varphi \right)^2 + \left(\partial_y \varphi \right)^2 \right]_{m\,\rm odd} = \frac{Lw}{2}\sum_{\substack{m\; \text{odd} \\n} } (q_x^2 + q_y^2 )\tilde\varphi(m,n) \left[\tilde\varphi(-m,n) +\tilde\varphi(-m,-n)\right]\;,
\end{equation}
where we used Eqs.~\eqref{app-eq:sum-x} and \eqref{app-eq:cos-all}. Similarly, the sum over the even modes becomes
\begin{equation}\label{app-eq:ham-even}
    \sum_{x=0}^{L-1}\sum_{y=0}^{w-1} \left[ \left(\partial_x \varphi \right)^2 +\left(\partial_y \varphi \right)^2 \right]_{m\,\rm even} = \frac{Lw}{2}\sum_{\substack{m\; \text{even} \\n} } (q_x^2 + q_y^2 )\tilde\varphi(m,n) \left[\tilde\varphi(-m,n) -\tilde\varphi(-m,-n)\right]\;.
\end{equation}
Finally, we use the mode relationship for pseudoscalars on a Klein bottle, Eq.~(\ref{app-eq:pseudoscalar-fourier}) to combine the sums~\eqref{app-eq:ham-odd}-\eqref{app-eq:ham-even} into
\begin{equation}\label{app-eq:hamiltonian-gauss}
    \mathcal{H} = \frac{\alpha L w}{2} \sum_{m, n} (q_x^2+q_y^2)\tilde\varphi(m,n)\tilde\varphi(-m, n) = \frac{\alpha L w}{2} \sum_{m, n} (q_x^2+q_y^2) |\tilde\varphi(m,n)|^2\;,
\end{equation}
where we used in the last step that $\varphi(-m,n) =\tilde\varphi(m,n)^*$. This final Fourier expansion shows the spin-wave Hamiltonian on the Klein bottle is still Gaussian. We will use this in the next section to compute spatial correlation functions.

\subsection{Spatial spin-correlations}
To verify the absence of long range order on the Klein bottle in the thermodynamic limit explicitly, we compute in the following the spatial correlations. Specifically, we seek an expression for the scaling of
\begin{equation}\label{app-eq:corr-gaussian}
    C^{\rm K}(\br,\br') = \langle \cos\left[ \varphi(\br) - \varphi(\br')\right]\rangle_{\rm K} =   {\rm e}^{- \frac{1}{2}\langle \left[ \varphi(\br)  - \varphi(\br')\right]^2 \rangle_{\rm K} },
\end{equation}
where we used that the Hamiltonian remains Gaussian on the Klein bottle to evaluate the average and include the subscript ${\rm K}$ to make explicit that we are performing computations on the Klein bottle. To evaluate Eq.~(\ref{app-eq:corr-gaussian}), we have to determine,
\begin{equation}\label{app-eq:angular-diffusion}
    \langle \left[ \varphi(\br)  - \varphi(\br')\right]^2 \rangle_{\rm K} = G^{\rm K}_\varphi(\br,\br) + G^{\rm K}_\varphi(\br',\br') - 2G^{\rm K}_\varphi(\br,\br'),
\end{equation}
where we define 
\begin{equation}
    G^{\rm K}_\varphi(\br,\br') = \langle \varphi (\br) \varphi(\br')\rangle_{\rm K}\;.
\end{equation}
To find $G^{\rm K}_\varphi(\br,\br')$, we use the Fourier decomposition given in Eq.~\eqref{app-eq:theta-fourier}. The average $\langle\tilde\varphi(m,n)\tilde\varphi(m',n')\rangle_{\rm K} $ reads
\begin{equation}
    \langle\tilde\varphi(m,n)\tilde\varphi(m',n')\rangle_{\rm K}  = \frac{k_b T }{2 \alpha L w( q_x^2 + q_y^2)} \delta_{m,-m'} \left[ \delta_{n,n'} - (-1)^m \delta_{n,-n'} \right]\;,
\end{equation}
where we used the properties of the Gaussian integral and the Kronecker delta's enforce the mode symmetry on a Klein bottle (on a torus one would simply have $\delta_{\bq,-\bq'}$). Straightforward algebra then leads to 
\begin{equation}\label{app-eq:angle-corr}
     G^{\rm K}_\varphi(\br,\br') = \langle \varphi (\br) \varphi(\br')\rangle_{\rm K} = \frac{k_bT}{\alpha L w} \left[ \sum_{\substack{m\; \text{odd} \\n} } \frac{{\rm e}^{iq_x(x-x')} \cos(q_y y ) \cos(q_y y') }{q_x^2 + q_y^2} +\sum_{\substack{m\; \text{even} \\n} } \frac{{\rm e}^{iq_x(x-x')} \sin(q_y y ) \sin(q_y y') }{q_x^2 + q_y^2}  \right]\;.
\end{equation}
As a result of the mode constraints imposed by the Klein bottle topology, $G^{\rm K}_\varphi(\br,\br')$ does not depend solely on $\br -\br'$, $i.e.$ $G^{\rm K}_\varphi(\br,\br') \neq G^{\rm K}_\varphi(\br -\br')$. However, Eq.~\eqref{app-eq:angle-corr} can still be conveniently written as 
\begin{align}
     G_\varphi^{\rm K}(\br,\br')  &= \frac{k_b T}{2 \alpha L w } \sum_{m,n} \frac{{\rm e}^{iq_x(x-x')}}{q_x^2 + q_y^2} \left\{ \cos\left[( q_y(y-y') \right] - (-1)^m \cos\left[ q_y(y+y') \right] \right\} \notag\\
     &= G_{\varphi}^{\rm T}(x-x', y-y') - G_{\varphi}^{\rm T}(x-x'+L, y+y'),\label{app-eq:angle-klein-torus}
\end{align}
where we used the known orientation correlation on a torus~\cite{goldenfeld2018lectures} defined over a toroidal double cover of the Klein bottle
\begin{equation}
    G_{\varphi}^{\rm T}(X, Y) = \frac{k_b T}{2\alpha L  w } \sum_{m,n} \frac{{\rm e}^{iq_x X}\cos(q_y Y)}{q_x^2 + q_y^2}.
\end{equation}
Equation~(\ref{app-eq:angle-klein-torus}) explicitly shows that -- due to the mode filtering on the Klein bottle -- $G^{\rm K}_{\varphi}(\br,\br')$ consists of decoupled contributions from torus-like correlations and correlations from the topological double-cover images (first and second term in Eq.~(\ref{app-eq:angle-klein-torus}), respectively). Finally, using the decomposition Eq.~\eqref{app-eq:angle-klein-torus} in Eq.~\eqref{app-eq:angular-diffusion} and fixing $y = y'=0$ and $x'=0$, spatial correlations on the Klein bottle computed from Eq.~\eqref{app-eq:corr-gaussian} read
\begin{equation}
    C^{\rm K}(\br, 0) \sim {\rm exp}\left\{ - \left[  G^{\rm T}_{\varphi}(0,0) -  G^{\rm T}_{\varphi}(x, 0) - G^{\rm T}_{\varphi}(L,0) + G^{\rm T}_{\varphi}(x+L, 0) \right] \right\} \;,
\end{equation}
which  tends in the limit $1\ll x \ll L$ to $C^{\rm K}(\br, 0)  \sim x^{-\eta}$ with $\eta = k_b T/(2 \pi \alpha)$, as in the case of a toroidal topology. Therefore, while the Klein bottle topology pins homogeneous orientation fields and bounds orientational fluctuations, global orientational order at finite temperatures is still solely quasi-long range. 

\begin{figure}[ht!]
    \centering
    \includegraphics[width=0.4\linewidth]{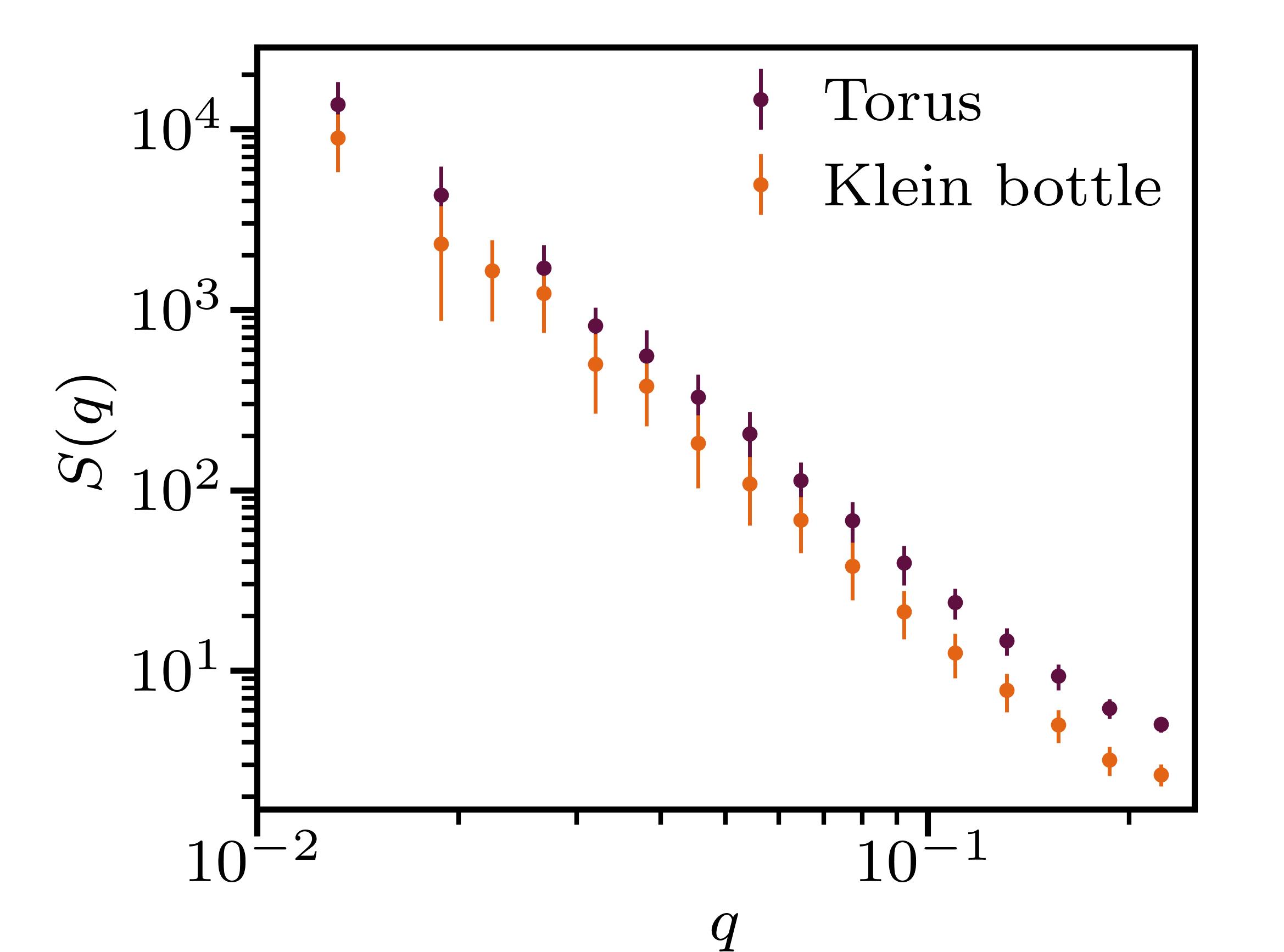}
    \caption{Numerical measurements of the angular average of the structure factor~(see Sec.~\ref{sec:structure}) for the Vicsek model on a torus (purple) and a Klein bottle (orange). Error bars indicate standard deviations measured over $N=24$ independent realizations.  Simulation parameters: $d_0 = 0.5$, $L=w=512$, $\eta=0.1$, $\rho_0=1$.}
    \label{smfig:fig1}
\end{figure}

\section{Structure factor measurements of the Vicsek model on a Klein bottle}\label{sec:structure}
The structure factor 
\begin{equation}\label{eq:sq}
    S(\bq) = \frac{1}{N} \langle   |\tilde\rho(\bq)|^2\rangle 
\end{equation}
determined from particle density distributions $\rho(\br)=\sum_i\delta(\mathbf{r}-\mathbf{r}_i)$ has been indicated as an effective way to numerically detect signatures of global directed motion in dynamical systems with ferromagnetic interactions~\cite{brambati2022signatures}. In the standard Vicsek model, $S(\bq)$ is scale-free~\cite{ginelli2016physics}, while it plateaus at large scales in the presence of an external field~\cite{brambati2022signatures}. To explore the analogy between domain topology and external fields discussed for passive systems in the main text [see Eq.~(\ref{eq:rescaling-B-K})], we thus measured the structure factor emerging from the Vicsek model dynamics on Klein bottle. In practice, we determine $S(\bq)$ given in Eq.~(\ref{eq:sq}) by averaging over independent realizations, and compute the Fourier-transformed density field as
\begin{equation}\label{app-eq:fourier-density}
    \tilde \rho(\bq) = \sum_{j=0}^{N} {\rm e}^{i \bq \cdot \br_j} + \sum_{j=0}^{N} {\rm e}^{i \bq \cdot \br_j'},
\end{equation}
where the first sum runs over the particles initialized on the original domain $\mathcal{D}$, while the second one runs over their topological image. The wave-vectors $\bq$ are taken from set of allowed modes for scalar fields on a torus ($\mathcal{Q}_{\rm T}$, see main text) and on a Klein bottle [$\mathcal{Q}_{\rm K}$ given in Eq.~(\ref{eq:QK}), subject to the mode exclusion defined by Eq.~(\ref{app-eq:scalar-fourier})]. We thus measure Eq.~\eqref{app-eq:fourier-density} in numerical simulations from which we compute Eq.~\eqref{eq:sq}, and then average its value over independent realizations. In addition, we perform an angular average and eventually analyze
\begin{equation}\label{app-eq:S(q)}
    S(q) = \langle S(\bq) \rangle_{|\bq| = q}\;.
\end{equation}
As shown in Fig.~\ref{smfig:fig1}, the structure factor behaves qualitatively in the same way as it does for the Vicsek model on a torus, \textit{i.e.} it does not show the plateauing behavior we would expect of the Klein bottle topology behaves in large systems like an external~field.

%

\bibliographystyle{apsrev4-2}
\bibliography{biblio}